\documentclass[10pt, conference]{IEEEtran}

\usepackage{times}

\usepackage{caption}
\usepackage[pdftex]{graphicx}
\usepackage{hyperref}
\usepackage{cite}

\newcommand{\citep}{\cite}

\begin{document}

\pagenumbering{gobble}

\title{\textbf{\Large Business Process Modeling and Execution--A Compiler for Distributed Microservices}}

\author{
\IEEEauthorblockN{Robert Singer}
\IEEEauthorblockA{
Department of Computer Sciences\\
FH JOANNEUM--University of Applied Sciences\\
Graz, Austria\\
robert.singer@fh-joanneum.at}
}

\maketitle

\begin{abstract}
In this paper, we propose to rethink the dominant logic of how to model business processes. We think that an actor based approach supports in a much better way the fundamental nature of business processes. We present a proposal for a compiler architecture to model and execute business processes as a set of communicating microservices that are hosted on a general purpose virtual machine for distributed execution.
\end{abstract}


\IEEEpeerreviewmaketitle

\hypersetup{pageanchor=false}

\section{Introduction and Motivation}
\label{introductionandmotivation}

Recent industrial activities demand initiatives towards the digital transformation of business processes. Furthermore, the industrial internet requires rethinking actual practices and tools for distributed business processes based on communicating human and (smart) machine actors.

During recent years, tools have emerged to support the execution of business processes, so-called Business Process Management Systems (BPMS). Most of these tools are build around the \emph{de facto} standard for business process modeling languages, namely Business Process Model and Notation (BPMN 2.x).

The standard may be suitable for modeling purposes, but does not directly support the execution of business process models~\citep{OMG:2013wz}~\citep{Silver.2011}~\citep{Borger:2011ib}~\citep{Singer:2014mz}. That means, there is a gap between the conceptual model and the digitized and executable representation. Furthermore, modeling of business processes needs analytical skills, experiences in abstraction and conceptualization using formal languages~\citep{Seymour:2012tv}~\citep{Seethamraju:2012en}, such as BPMN. These capabilities are typically not available in many companies, especially not in Small and Medium Enterprises (SME)~\citep{Singer:2015hl}.

To overcome this weakness, modeling notations based on actor models have emerged. The standard for an actor based approach for business process modeling is the so-called Subject-oriented Business Process Management (S-BPM)~\citep{Fleischmann:2012va} approach.

In summary, S-BPM treats any business process as a (loosely) coupled network of Actors, which can be seen as microservices. S-BPM is a mature approach, as has been proven in theory and practice~\citep{Fleischmann:1994wf}~\citep{Fleischmann:2012va}~\citep{Rass:2013vn}~\citep{Fleischmann:2015}~\citep{metasonic}. Nevertheless, in this work we now want to propose a further paradigm change and to get rid of monolithic applications for the execution and management of business processes. 

In this work, we investigate how to compile directly S-BPM models into a set of executable processes or microservices that coordinate work through the exchange of messages. We discuss general concepts for such an approach and present an architecture for a concrete realization dependent on certain technological decisions.

In \autoref{traditionalworkflowsystem} we review the typical architecture of workflow systems. Then in \autoref{actorbasedworkflowsystems} we present and discuss an architecture of actor based workflow systems. Based on this we will work out the main differences between those two methodologies. Furthermore, after a short literature review of a ``compiler-engine'' proposal in \autoref{businessprocesscompiler} we will propose and discuss an approach based on actors. Furthermore, we give a short review of S-BPM. This includes the explanation of the modeling notation and the formal background behind the scenes. Afterward, we discuss a methodology to translate S-BPM models into compilable code.

\section{Traditional Workflow System}
\label{traditionalworkflowsystem}

\begin{figure}[htbp]
\centering
\includegraphics[keepaspectratio,width=8.5cm,height=0.75\textheight]{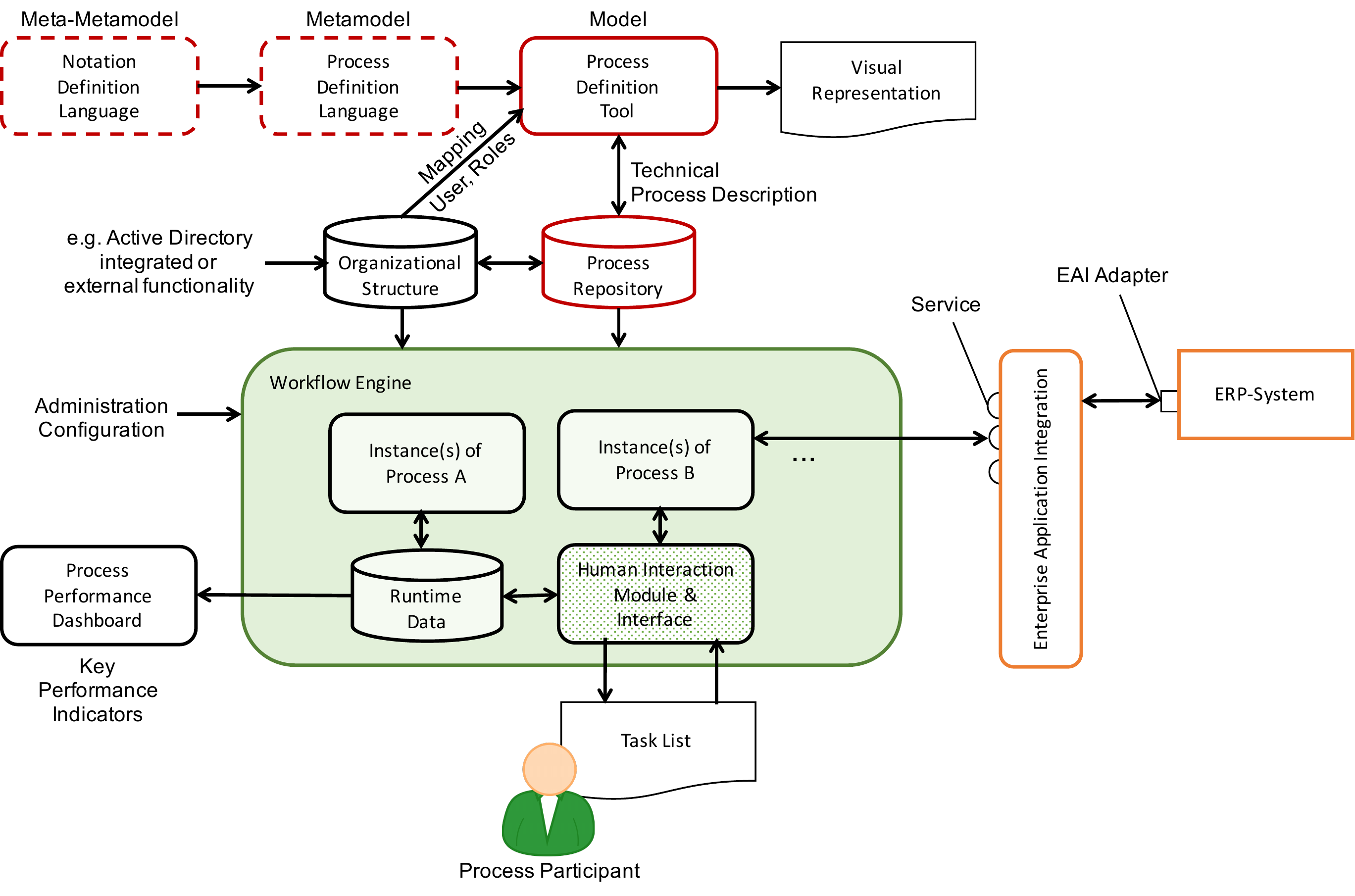}
\caption{Typical architecture of a workflow system.~\citep{Singer:2016xy}}
\label{fig-1}
\end{figure}

A typical conceptual architecture of a workflow system as part of a business process management system is depicted in \autoref{fig-1}.

A short description of the depicted architecture is as follows: business process models are stored in a repository; we also need a formal model of the organization, so we can link organizational groups, roles and individual persons with the activities of the process model (who is doing what). Any process also has transient entities, which change state during process execution---the so-called business objects (BO). Process models can be uploaded to and started by the workflow engine; they are interpreted by the software logic of the application (process execution). Furthermore, it is important to understand that there must be some mechanism to interact with human process participants. Firstly, the application needs to create and maintain a task list to distribute work~\citep{Russell:2004ix}. Secondly, a task requires some input (data) from a human actor. Typically, a task is presented as a form based window to the process participant, which includes some read-only data and offers some interactive elements to enter or change data; the forms have to be designed and developed manually, or, can be generated automatically from business objects. Finally, we also have the need to interact with other systems in the enterprise, which nowadays typically is done via service calls. 

The challenge lies in the automatic translation from process model to executable model. That is the reason the BPMN 2.x standard document differentiates between several compliance classes. 

However, the dominant logic how to define business processes is not necessarily the only thinkable one and other approaches offer some promising possibilities to overcome several of the obstacles we are confronted with in the practical use of BPMS.

\section{Actor Based Workflow Systems}
\label{actorbasedworkflowsystems}

\subsection{Motivation}
\label{motivation}

We agree that software engineering has to support business and not the other way round. However, it could be prolific to learn from software engineering~\citep{gruhn2007}. Over the last couple of years, new concepts have emerged or got more attention, as, reactive and flow-based programming. There is also an increasing interest in microservices, functional programming and actor based systems to support the need to develop solutions that are responsive, resilient, elastic and message driven---the core requirements stated in the Reactive Manifesto~\citep{boner}. 

Furthermore, large systems are composed of smaller ones and, therefore, depend on the reactive properties of their constituents. This leads to the concept of microservices~\citep{Newman:2015ye} as a design pattern to build reactive systems meeting the requirements mentioned before. The microservice architectural style is an approach to understanding any application as a collection of small services, each of them running in its process environment and communicating with lightweight mechanisms; state changes of a service can only be triggered by receiving certain message types and business objects (data).

\subsection{Foundation}
\label{foundation}

The proposal now is, that based on formal modeling approaches rooted in computer science---as mentioned before---it is possible to derive innovative methods to better model business processes. We think, that business process models based on actor models~\citep{Hewitt:1973xq} are a premium candidate for new approaches in the domain of BPM; the actor model supports all mentioned requirements. 

A nearly similar approach has been proposed by Albert Fleischmann who developed a business process modeling methodology called PASS Parallel Activity Specification Scheme (PASS)~\citep{Fleischmann:1994wf}, which itself is based on Calculus of Communicating Systems (CCS)~\citep{Milner:1980qr}~\citep{Aitenbichler:2011lr}~\citep{Borgert.2011}, a process calculus describing reactive systems~\citep{Aceto.2007}. Later, this concept has evolved into the so-called Subject-oriented Business Process Management (S-BPM) methodology~\citep{Fleischmann:2012va}. In short, this methodology includes all necessary concepts to define reactive and executable models of business processes.

An S-BPM process is defined via the communication exchange channels between subjects (actors are instantiated subjects in this context, or the other way round---subjects are generalizations of actors). Additionally, each subject has a defined (but invisible to the outside world) internal behavior, which is determined as a process flow using states for receiving or sending a message (to another subject), and states in which the subject is doing some work. States can be flagged as starting or ending states and are connected using directed arcs. In our approach, we think of subjects as actors. In the context of BPM, actors define who is doing what, as they are mapped to a resource for execution (organizational roles). Typically, S-BPM models consist of two types of representations: a Subject Interaction Diagram (SID) and a set of Subject Behavior Diagrams (SBD). The SID includes the subjects (this are the actors), the messages exchanged between the actors and the business objects attached to the messages. The SBD includes all possible state sequences of an actor: a finite set of send, receive and function states. 

\begin{figure}[htbp]
\centering
\includegraphics[keepaspectratio,width=8.5cm,height=0.75\textheight]{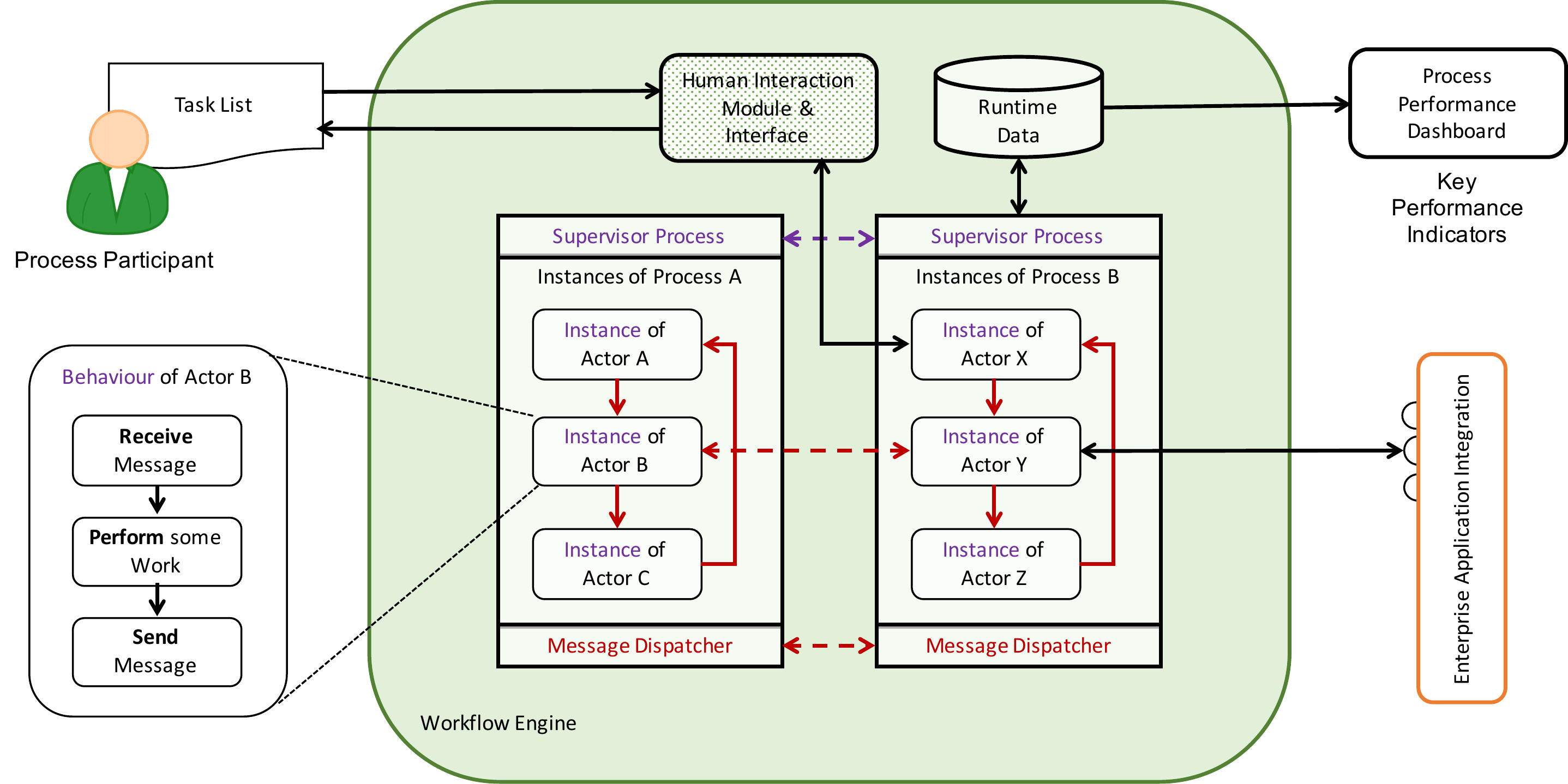}
\caption{Architecture of an actor based workflow system.~\citep{Singer:2016xy}}
\label{fig-2}
\end{figure}

\begin{figure}[htbp]
\centering
\includegraphics[keepaspectratio,width=8.5cm,height=0.75\textheight]{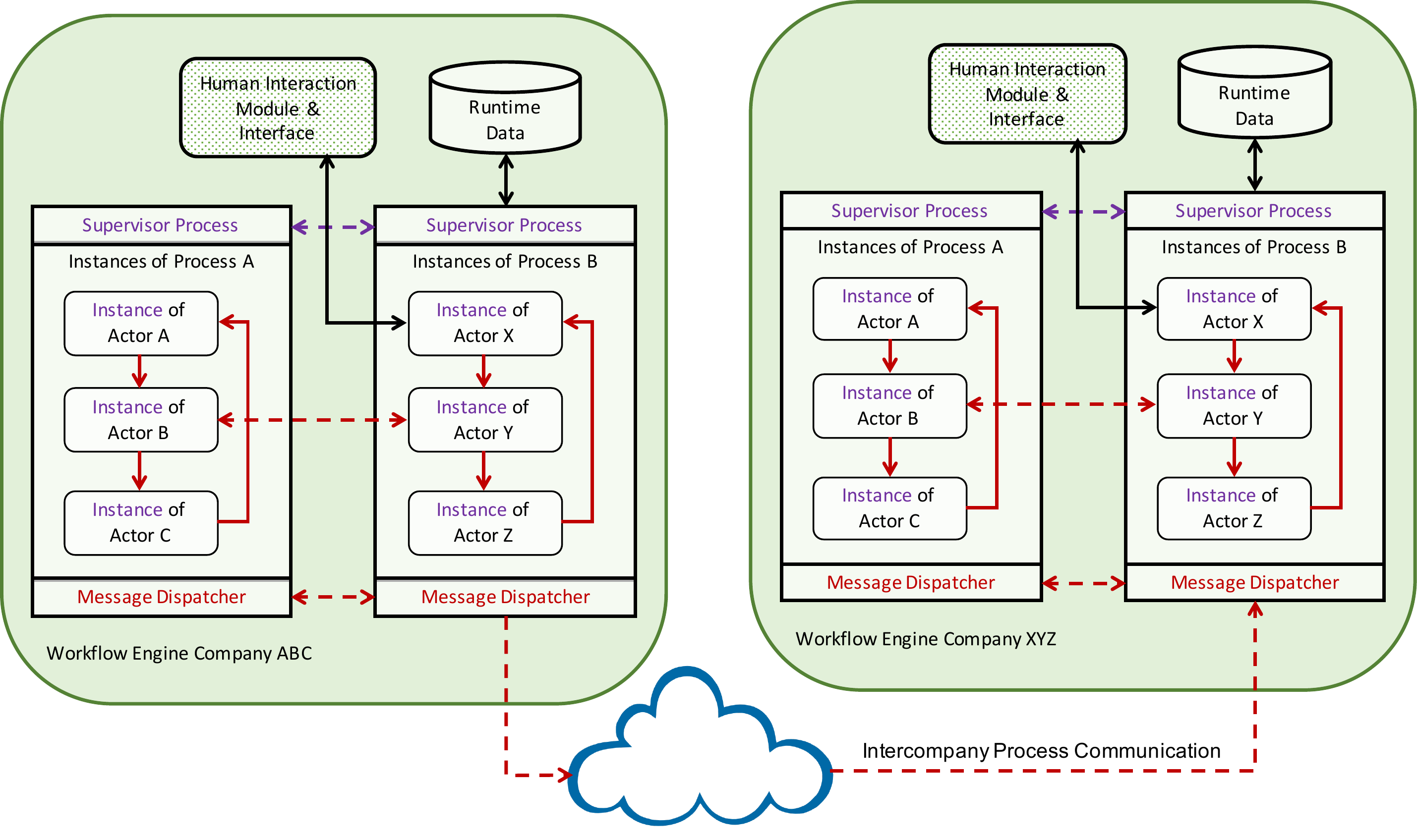}
\caption{Architecture of an actor based and distributed workflow system.~\citep{Singer:2016xy}}
\label{fig-3}
\end{figure}

According to our definition, any actor can be represented as a Finite State Machine (FSM) with states as mentioned above, the so-called internal behavior. Furthermore, state changes can be triggered by receiving messages from other actors~\citep{Brand:1983vx}. An actor or subject is a general concept and can be instantiated by a human or machine. Based on the following conceptual architecture, as depicted in \autoref{fig-2} and \autoref{fig-3}, we have developed an entirely functional S-BPM workflow solution using the Microsoft Windows Workflow Foundation functionality as discussed in~\citep{Singer:2014} and~\citep{Singer:2015}; for a more detailed discussion of the architecture please consult these references.

A generalized architecture of an actor based workflow systems is schematically depicted in \autoref{fig-2}. 

The main components of the actor-based architecture are as follows (\autoref{fig-2}): the workflow engine manages the business processes via supervisor processes; the workflow engine loads the model from the repository and starts the supervisor process that itself is responsible for starting all needed actors. Each supervisor process manages one instance of a process model. When a business process ends, the responsible supervisor process is terminated. Each business process has a dedicated instance of a \emph{message dispatcher} that is responsible for routing the messages between all actors and between actors and the supervisor process (\emph{intra-process} communication). Sending messages between \emph{message dispatchers} establishes \emph{inter-process} communication.

In \autoref{fig-3} we see, that based on the depicted architecture it is possible to distribute actor based business processes over more than one workflow engine, as long as we can route messages between them. Routing messages between actors is a significant step to establish \emph{inter-company} business processes.

\section{Business Process Compiler}
\label{businessprocesscompiler}

\subsection{Compiler-Engine Architecture}
\label{compiler-enginearchitecture}

It seems worth to think about \emph{compiling} business process models into executable code. Recently this has been discussed by~\citep{Prinz:2015} who proposed a compiler-engine architecture.

One of the core concepts of this conceptual architecture is the proposal to use a virtual machine for the execution of business processes, as also discussed in~\citep{Prinz:2015uz}. Such a virtual machine should be independent from the notation used to model a process. As logical consequence there is the need for an intermediate representation (IR), which cannot be directly used for modeling: 

\begin{quote}

It is necessary to provide a more high-level but IR-conform processing language (like a subset of BPMN and EPC). Therefore, that language has to be automatically transformable into the IR. 
\end{quote}

In the first step of the translation process a parser creates a parse tree, a Process Structure Tree (PST), which is then used for translation; this is conform with compiler theory. For a further discussion of this approach we refer to the work of Prinz. et al. ~\citep{Prinz:2015}~\citep{Prinz:2015uz}.

\subsection{An Actor-based Approach}
\label{anactor-basedapproach}

Based on similar ideas, but motivated by the actor model, we propose to investigate a different approach. Thinking bottom up, we can review available technology which inherently supports the development and execution of actor models. Because of the underlying formal models, we think that a functional programming language would be a good starting point for further research. The main reason is that business processes are highly concurrent; the lack of mutable state---as in functional languages---makes concurrency almost trivial.

\subsubsection{A Virtual Machine for Processes}
\label{avirtualmachineforprocesses}

So we could easily identify, for example, the following frameworks and\slash or programming languages as candidates for hosting business processes modeled as communicating actors: AKKA and Scala (on the Java Virtual Machine) and Erlang or Elixir (on the Erlang Virtual Machine)

Using a Virtual Machine (VM) has many advantages, such as, for example, to provide an abstraction layer from the operating system. Furthermore, we think that it is much more efficient using a well established and general purpose VM instead developing a specific BPM-VM---as long as it fits the purpose. Verifying the feasibility of using the Erlang Run-Time System (ERTS) as VM for the execution of compiled business process models is the aim of this research proposal. 

People (and machines) coordinate work through the exchange of messages. This is our understanding of what a business process is---and that is the essence of Erlang; consequently, we see business processes as a network of connected microservices. This sounds technical, but a microservice can also be realized by a human actor. A microservice in our setting is simply an Erlang process.

In Erlang we have a hierarchy of (linked) processes that exchange messages; these message exchanges can also be done between processes hosted on different VMs that again can be hosted on different computers that are located in different networks. Furthermore, the language capabilities of Erlang fit perfectly with the modeling concept of Subject-oriented Business Process Management.

It is not our intention to propose the development of another workflow engine, but to develop and study an software architecture to compile process models based on the S-BPM methodology into executable code, which furthermore can be executed on the chosen generic Virtual Machine.

\subsubsection{S-BPM as Modeling Language}
\label{s-bpmasmodelinglanguage}

\begin{figure}[htbp]
\centering
\includegraphics[keepaspectratio,width=8.5cm,height=0.75\textheight]{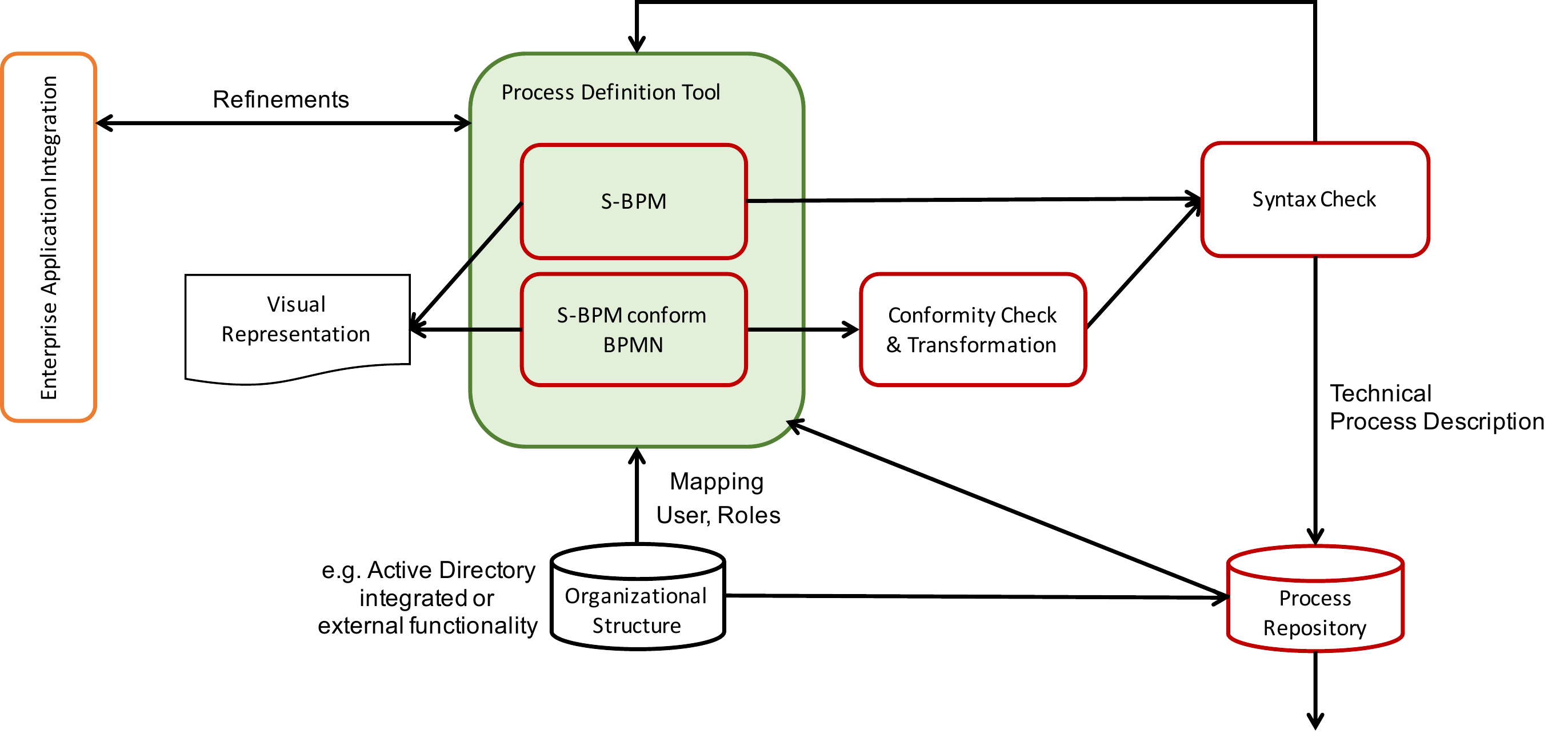}
\caption{Creating a S-BPM process model with a Process Definition Tool: the S-BPM model can be defined graphically, as formal language sentences~\citep{Fleischmann:2012va}~\citep{Langmann:2014qa}, or with a reduced set of BPMN 2.x symbols~\citep{Albert:1} as long as the serialized model conforms with an XML-schema definition for the tool (BPMS) that uses the XML-file for execution.~\citep{Singer:2016xy}}
\label{fig-6}
\end{figure}

All relevant modeling notations (BPMN, eEPC, or S-BPM) for business processes are based on a graphical representation. The semantic of the models is typically serialized as some XML structure. Such an XML-file can be used as input for the compilation process as depicted in \autoref{fig-6}.

As noted in \autoref{fig-6} it is possible to show, that we can use BPMN with a restricted set of modeling elements to define S-BPM processes models. In that case, we have to transform the BPMN-XML into the form of the used tool. As with any BPMS, the organizational structure has to be mapped to the process. The processes are stored in the process repository. \autoref{fig-6} is a modified version of the upper part of \autoref{fig-1}. Function states can include Refinements that means calls of external services. 

S-BPM is a fully quantified and formal language to define the interaction of actors and to describe their internal behavior in a heterogeneous (human and machine actors) multi-actor environment.

\subsubsection{From Model to Code}
\label{frommodeltocode}

\begin{figure}[htbp]
\centering
\includegraphics[keepaspectratio,width=8.5cm,height=0.75\textheight]{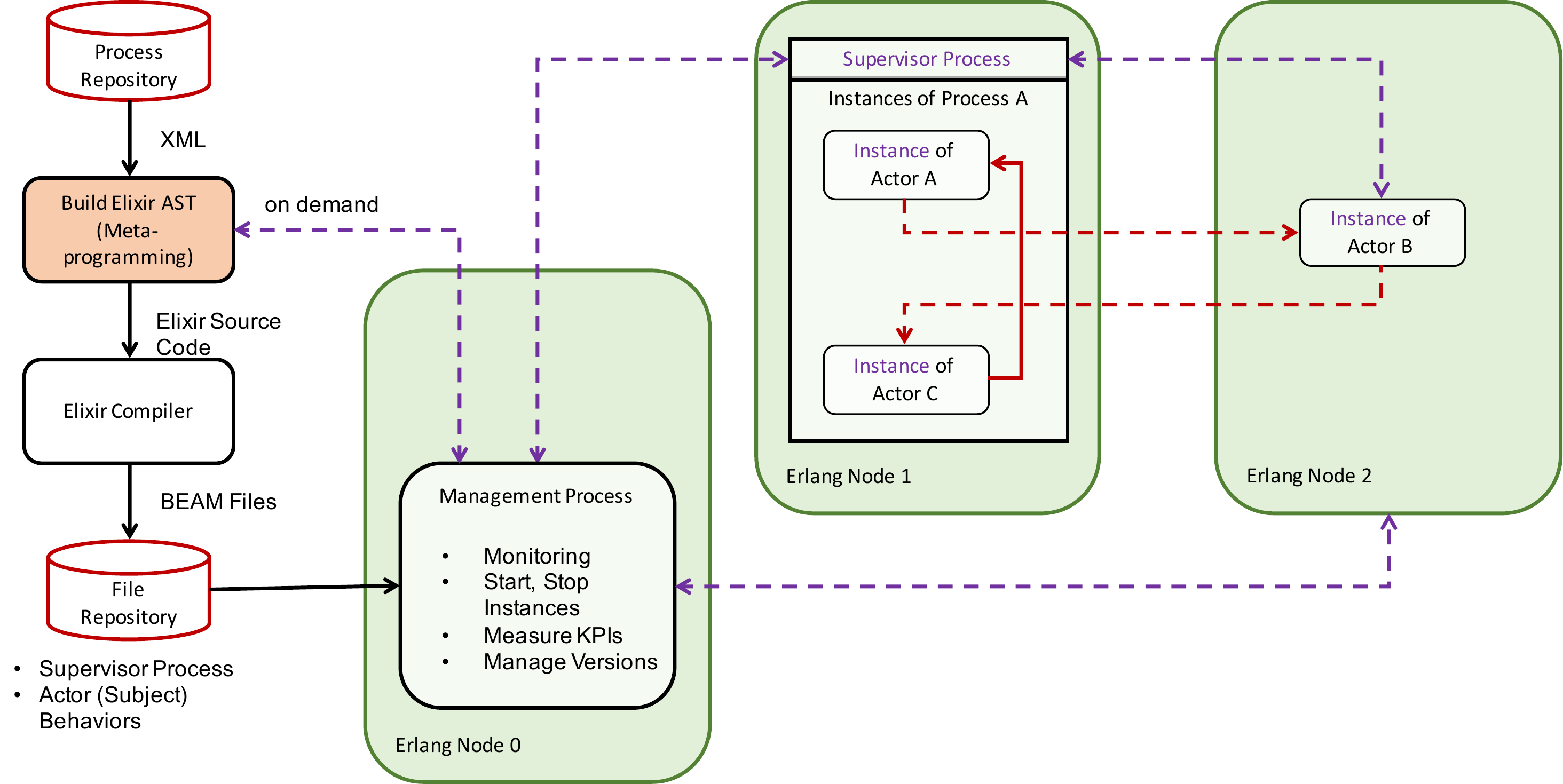}
\caption{Consumer side: the model is translated into executable code. The compiled processes are loaded and supervised by a management process. Processes can be executed on different Erlang nodes, even running in distinct networks.~\citep{Singer:2016xy}}
\label{fig-7}
\end{figure}

On the consumer side (the side that uses the models) the models are compiled into so-called .beam files. Each subject has to be converted into an Erlang process and can be executed on any Erlang node (see \autoref{fig-7}). A supervisor process is linked with each subject to control each instance and the network of communicating subjects---the whole business process. This is also a standard approach for Erlang applications; please consult ~\citep{Armstrong:2013qy}~\citep{Cesarini:2015zr}, for example, for more information about Erlang. 

The idea of how to compile S-BPM processes builds on the concept of metaprogramming---code that writes code. That means, the process model, consisting of several communicating subjects is transformed via a translation process into a source code file. This could be Erlang code, but as Elixir~\citep{Thomas:2016} offers specific support for metaprogramming~\citep{McCord:2015ul} we have chosen it as our choice for the transformation process. Elixir builds on the Erlang VM and source code is also compiled into .beam files. Elixir source code can easily be mixed with Erlang code.

Elixir code is represented internally by the abstract syntax tree (AST). Most languages have an AST, but it is typically not visible and accessible to the programmer. When programs are compiled or interpreted, their source is transformed into a tree structure being turned into byte-code or machine-code. This process is usually masked away. In Elixir the AST is exposed in a form that can be represented by Elixir's own data structures and a natural syntax to interact with it. Therefore, code can use the language capabilities to write directly code; code interprets the model and constructs a source file that represents the model as Elixir source file.

\begin{figure}[htbp]
\centering
\includegraphics[keepaspectratio,width=8.5cm,height=0.75\textheight]{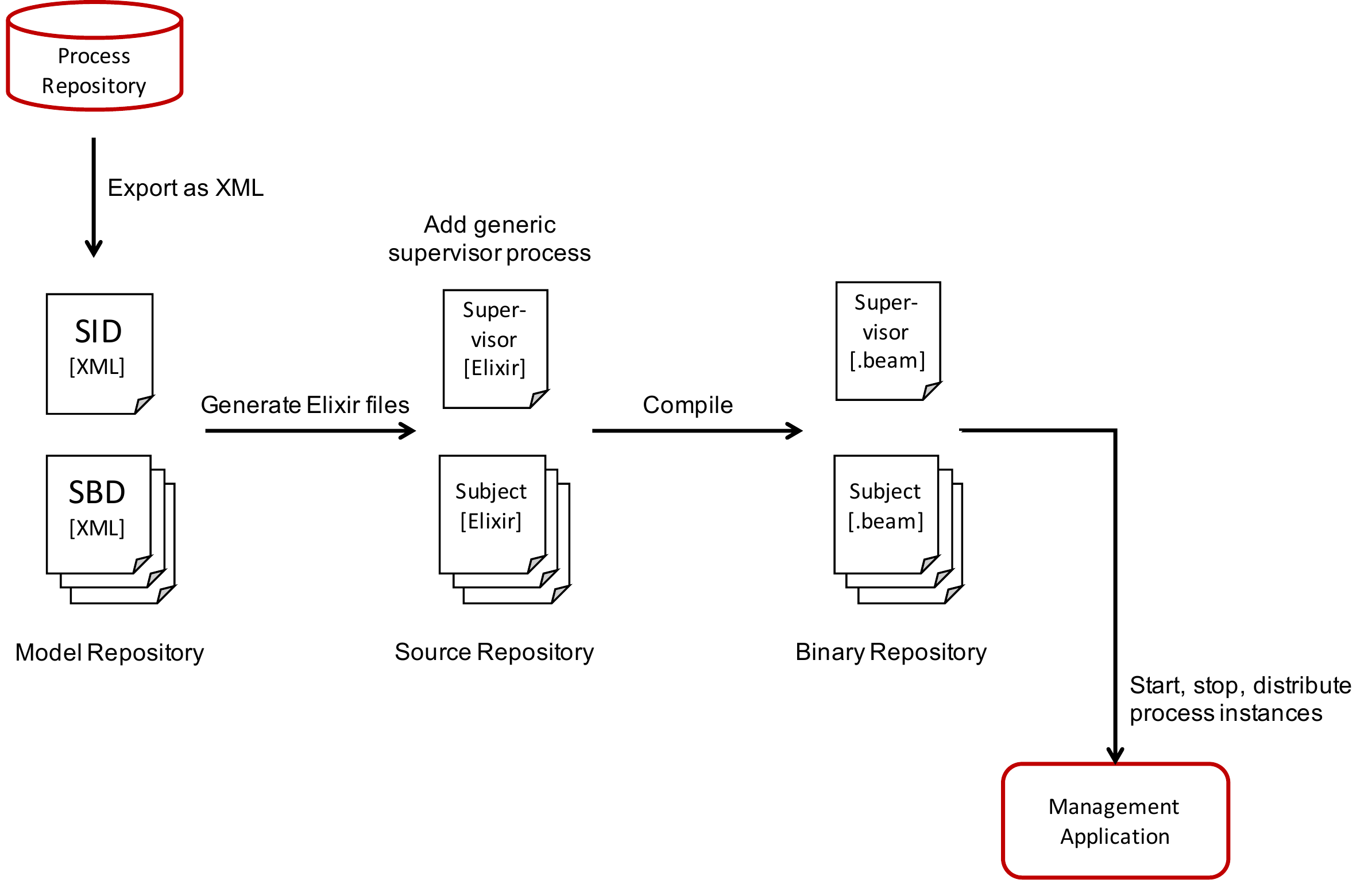}
\caption{The compilation work flow.~\citep{Singer:2016xy}}
\label{fig-18}
\end{figure}

The full compilation (production) flow is depicted in \autoref{fig-18}. First the process models are exported from the repository in some XML format. This leads to a set of files: One file for the Subject Interaction Diagram (SID) which defines the actors (subjects) of the business process, the messages between all subjects and the business objects attached with the messages; business objects can be simple or complex data structures. For each Subject Behavior Diagram (SBD) we get a corresponding file defining the internal behavior of the actor (subject). 

Now, a parser takes these files as input and generates for each SBD an Elixir source file. The information contained in the SID-file is used to produce the message exchange behavior of each SBD-file. Furthermore, based on a provided template file, a supervisor process for the specific business process has to be generated. The supervisor process has the ``control'' over all other processes; that means this process is informed, for example, if one of the subject instances crashes, or it can be designed for performance measures (to measure a service level, for instance). It can also be used for applying business rules to the overall message exchange between individual subjects~\citep{Singer2015}. 

The generated Elixir source files are compiled with the Elixir compiler and are stored in a repository. An application has to be provided to manage the set of executable binaries and the running instances. So it has to provide functionality to start a business process on one or more Erlang nodes---the virtual machines. Furthermore, as supported by the ERTS, this application could also update running instances with a new version (of course, this has to consider the actual state of the process or process component).

\section{Conclusion and Future Work}
\label{conclusionandfuturework}

The feasibility to model any S-BPM process as Elixir or Erlang process has been proven based on hand-crafted code snippets; nevertheless, this is a trivial fact and only based on the fact that both S-BPM and Elixir are based on communicating actors. However, how exactly to define a suitable data model for S-BPM processes and how to automatically generate source files from this data model (via metaprogramming) has to be investigated in detail and is an ongoing research activity.

For practical use, further aspects have to be considered. We have identified the following topics for which solutions have to be identified and integrated into an overall Business Process Management System (BPMS). This is also under development; our research is dedicated for industrial use, and many topics have to be investigated:

Firstly, there must also be a mechanism for the distribution of work for human actors; for example, there is no guarantee, that a task will be started or finished as intended. How to handle all these possible and thinkable exceptions in a human interaction workflow?

Another topic is the integration into an organizational structure (roles and access rights), which can be challenging in the case of distributed processes. There are some ideas how to handle this issue, but they have to be investigated in detail.

The most critical and yet not fully understood problem seems to be security. The ERTS provides a mechanism to control the rights to \emph{run} a process, but there appears to be the need for more research to understand fully the implications for the execution of business processes---especially if processes span over more than one organizational unit.

A concept to handle business objects has to be developed: in a technical sense, how to send business objects from an actor in company A to an actor in company B---considering rights and safety issues, data integrity and so forth.

Furthermore, during development and research, other approaches emerged that are also currently under investigation: we call it smart or intelligent business objects (smartBO). A business process is based on data, the business objects. Often BOs are seen as second class entities in a classical view. In our concept, smartBO know the business process. That means, dependent on their state; they know who is the next actor according to the process definition and they forward themselves to the next actor (or actors) for processing. 

Up to now the discussed approach and design decisions have been proved very fruitful. Nevertheless, a lot of research and development work has to be done.

\bibliographystyle{IEEEtran}
\bibliography{IEEEabrv,iara}

\end{document}